\documentclass[12pt]{article} 
\usepackage{epsf}

\begin{document}
%\mbox{}
\begin{center}
{\bf \Large Upsilon ($\Upsilon$) Dissociation in Quark-Gluon Plasma}
\end{center}
\vspace*{1cm}
\begin{center}
{\bf Sidi Cherkawi Benzahra\footnote[1]{benzahra@physics.spa.umn.edu}}
\end{center}
\begin{center}
{\large School of Physics and Astronomy}
\end{center}
\begin{center}
{\large University of Minnesota}
\end{center}
\begin{center}
{\large Minneapolis, MN 55455}
\end{center}
\vspace*{1cm}
\begin{abstract}
\noindent
I consider the dissociation of $\Upsilon$ due to absorption of a thermal 
gluon.  I discuss the dissociation rate in terms of the energy
density, the number density, and the temperature of the quark-gluon plasma.  
I compare this to the effect due to screening. 
\end{abstract}

\hspace{0.3in} When the bottom quark or anti-quark is struck by a
high energy gluon, the upsilon meson can dissociate into other elements.  The 
medium, the quark-gluon plasma, can be full of gluons that can cause this 
dissociation, and this can happen by exciting color-singlet 
${\mid b\bar{b} \rangle}^{(1)}$ into a
color-octet continuum state. The bottom quark absorbs energy from the
gluon field.  When the bottom quark and anti-quark are close
to each other, asymptotic freedom comes into play, and the binding energy
can be derived the same way we derive it for the hydrogen atom. To an 
approximation, there is a parallelism between the two physics [1].  For the singlet 
state the energy is:

\begin{equation}
{E}={4\over9} \alpha_{s}^2m_{Q} \, ,
\end{equation}

and the radius is:
\begin{equation}
{a}={{3}\over{2{\alpha_{s}}m_{Q}}} \, . 
\end{equation}

The wavelength of the gluon that dissociates this state fits the radius of
the singlet state of $\Upsilon$ to a good approximation. 
The S matrix in this case is:
\begin{equation}
{S_{fi}}=-i\int_{-\infty}^ {\infty} dt \langle {\rm octet} \mid grE^{a}\cos\theta
\mid 
 {\rm singlet} \rangle 
\end{equation}
where $\rm E^{a}$ is the color electric field. Taking the singlet state to be a
1s-wavefunction and the octet to be a plane wave, and using the dipole
moment matrix, the calculation of the S
matrix in terms of the relative momentum of the $ {\mid b\bar{b}
\rangle}^{(8)}$
pair can be tedious, but straightforward. The ionization of the hydrogen atom 
by electromagnetic radiation leads to a similar S matrix [1,2].
\begin{equation}
{S_{fi}}={{32\pi gE^{a}(\omega)ka^{5}\cos\theta}\over{\sqrt{6}\sqrt{\pi
a^{3}V}{(1+k^{2}a^{2})}^{3}}} \, . 
\end{equation}

Here $\omega$ is 
the difference between the octet state energy and the
singlet state energy, k is the relative momentum of the bottom quark in the octet
state, a is the radius of the singlet state, and V is the
quantization volume.
Assuming color neutrality of the medium
\begin{equation}
\langle E_{i}^{a}(\omega) E_{j}^{b}(\omega) \rangle={{1}\over{24}} \delta_{ij}
\delta_{ab} \langle {\mid E(\omega)
\mid}^{2} \rangle \, .
\end{equation}
Using Fermi's Golden Rule, the transition rate is:
\begin{equation}
R_{fi}={2\pi} \rho(k) {\mid \langle {\rm octet} \mid g r E \cos\theta \mid  
{\rm singlet} \rangle \mid}^{2} 
\end{equation}
where
\begin{equation}
{\rho(k)}={{m_{Q}Vk\sin\theta d\theta d\phi}\over{8 {\pi}^3}} \, . 
\end{equation}
Integrating over k, the transition probability becomes
\begin{equation}
{P_{fi}}={2\over{3}} \pi \alpha_{s} a^{2} \langle {\mid E(\omega)
\mid}^{2} \rangle  \, .
\end{equation}

There is a threshold energy of about 850 MeV for the dissociation of the 
upsilon meson into two highly-energetic bottom quarks.  Only gluons with energy 
exceeding $\omega_{min}$= 850 MeV can dissociate
the upsilon.  So the relevant energy density is not just the average energy
density, but the energy of the gluons
which have an energy higher than the threshold energy.  In deconfined
matter, such as quark-gluon plasma, we expect gluons in a medium of
200 MeV temperature to have an average momentum of 600 MeV [3].  So
there are some higher-energy gluons which overcome the 850 MeV and dissociate
the upsilon.\\

Dividing both sides of eq.(8) by the total interaction time, $\tau_{fi}$, I get
\begin{equation}
\Gamma_{\rm dis}={2\over3}\pi a^{2} \alpha_{s} {{\langle {\mid E(\omega)
\mid}^{2}\rangle}\over \tau_{fi}} \, , 
\end{equation}
where ${{\langle {\mid E(\omega)
\mid}^{2}\rangle}/ \tau_{fi}}$ is the color-electric power density of the
medium, which can be evaluated analytically for a variety of models. One such
model is a dilute gas of color charges. Denoting the Casimir of the color 
charge by $Q^{2}$ in this model, it is found that
\begin{equation}
{{{\langle {\mid E(\omega)\mid}^{2}\rangle}\over \tau_{fi}}}\approx
{{\pi\over{2}} \alpha_{s}Q^{2}\tilde{\rho}(w)} \, ,
\end{equation}

where $\tilde{\rho}(w)$ is the weighted average of the density of charges in the medium.
Combining eqs.(9) and (10), and introducing the number density of gluons, the
dissociation rate of the upsilon meson becomes [1]:
\begin{equation}
\Gamma_{dis}\approx {8\over9}\pi^{3} {\alpha_{s}}^2 a^2 n \, . 
\end{equation}

This dissociation rate can be calculated in terms of the temperature of the
quark-gluon plasma. But I will only include the gluons with energy exceeding 
the threshold energy of dissociation, $\omega_{min}$.  For a medium of gluons
\begin{equation}
n=N/V={1\over{2\pi^2}} \int_{\omega_{min}}^{\infty} {d\omega} {{\omega^2}\over 
{{\rm exp}{(\omega/T})}
 -1} \, ,
\end{equation}

which gives us
\begin{equation}
\tau_{dis} \approx {{{m_{Q}^{2}}}\over{\pi
\sum_{k=1}^{\infty} 
\left[ \left ({{T}\over{k}} \right )\omega^2_{\rm min}
+2{\left ({{T}\over{k}}\right )^2}
\omega_{\rm min}+ 2 \left ({{T}\over{k}}\right )^3 \right ]
e^{-k\omega_{\rm min}/T}}} \, .
\end{equation}

The minimum temperature required to achieve deconfinement is generally
understood to be about 150 to 200 MeV.  RHIC, the Relativistic Heavy Ion 
Collider, is the first collider designed to specifically create this plasma.  It
may reach a temperature of 500 MeV.  CERN, on the other hand,  hopes to reach a 
temperature of 1 GeV by colliding heavy nuclei in the Large Hadron Collider.  
Inserting a temperature of 500 MeV into eq.(13) I get a dissociation time 
of 4 fm/c, which is comparable to the life span of the
quark-gluon plasma--a typical life span of a
quark-gluon plasma is about 2 to 5 fm/c [4].  If I use the CERN 1 GeV 
temperature, I get $\tau_{\rm dis}\approx 0.55\; \mbox{fm/c}$.  
See figure 1.\\

In addition to the dissociation of upsilon by gluons, there is another
kind of dissociation which is caused by the screening of the color
charges of the quarks in the medium [5].  In the high temperature deconfined 
phase, the bottom-antibottom 
free energy $V_{b\bar{b}}$, which is the Debye potential with inverse screening 
length $m_{el}$, is given by [6]

\begin{equation}
{V_{b\bar{b}}}=-{4 \over {3}}{{\alpha_{s}} \over {r}}e^{-m_{el}r} \, .
\end{equation}
Using a variational calculation with an exponential trial wavefunction,
$Ae^{-r/a_{T}}$, I look for a critical value of $m_{el}$ where the
upsilon meson is no longer bound. I find this critical $m_{el}$ to be
\begin{equation}
{m_{el}}= {2 \over {3}} \alpha_{s} m_{Q} \, .
\end{equation}
Here I use [7] $\alpha_{s}(m_{b}) = 0.2325 \pm 0.0044$.  Inserting $m_{el}$ 
in the equation [6]
\begin{equation}
{m_{el}^{2}} = {1 \over {3}} g^{2} (N + {{N_{f}} \over {2}}) T^{2} \, ,
\end{equation}
where N=3 from the SU(N) group and $N_{f}=3$ is the number of light flavors, 
and using
the temperature-dependent coupling constant as given by [6]
\begin{equation}
{{g^{2}} \over {4 \pi}} = {{6 \pi} \over {27 \rm ln \left(T/50 \rm MeV \right)}}
\end{equation}

I find that the ground state of upsilon is unbound at a temperature of 
T=250 MeV. Above this temperature, the effect of screening does not allow 
the upsilon meson to exist in a 1s state.\\

According to the above results, the time of dissociation of the upsilon meson 
seems to be comparable to the life span of the quark-gluon plasma. If we 
calculate the dissociation time for 
$J/\psi$, using the same principle we used above, we will find it to be even less 
than the one for the $\Upsilon$.  This works better for the $J/\Psi$ meson 
because its quark, c, is light compared to the b quark of the $\Upsilon$, and 
also the binding energy of the $J/\Psi$ is smaller compared to the binding 
energy of the $\Upsilon$. Screening plays a major role when the temperature
of the quark-gluon plasma exceeds 250 MeV. If there is a suppression of upsilon
mesons, it is more likely due to the screening effect than it is to the high
energy gluon absoption or collision.  I conclude that high energy
gluons and mostly the effect of screening are indispensable for the dissociation 
of the upsilon meson in the quark-gluon plasma. \\

\begin{center}
{\bf \large Acknowledgment}
\end{center}
I am indebted to Benjamin Bayman, Mohamed Belkacem, Paul Ellis, Joseph Kapusta, and Stephen Wong for
their help and advice on this paper.\\

\begin{center}
{\bf \large References}
\end{center}
[1] B. Muller, preprint nuc-th/9806023 v2 14 July, 1998(unpublished) .\newline
[2] L. I. Schiff, Quantum Mechanics (McGraw-Hill, New York, 1968).
\newline
[3] D. Kharzeev and H. Satz, Physics Letters B 334, 155 (1994). \newline
[4] S. A. Bass et. al., preprint nucl-th/9902055 Feb. 22, 1999\newline 
[5] T. Matsui and H. Satz, Physics Letters B 178, 416 (1986). \newline
[6] J. I. Kapusta, Finite-Temperature Field Theory (Cambridge University 
Press, Cambridge, 1989). \newline
[7] Matthias Jamin and Antonio Pich, preprint hep-ph/9702276 Feb. 6, 1997. \newline

\begin{figure}[h]
\vskip3mm
\epsfxsize=10cm
\center{\epsfbox{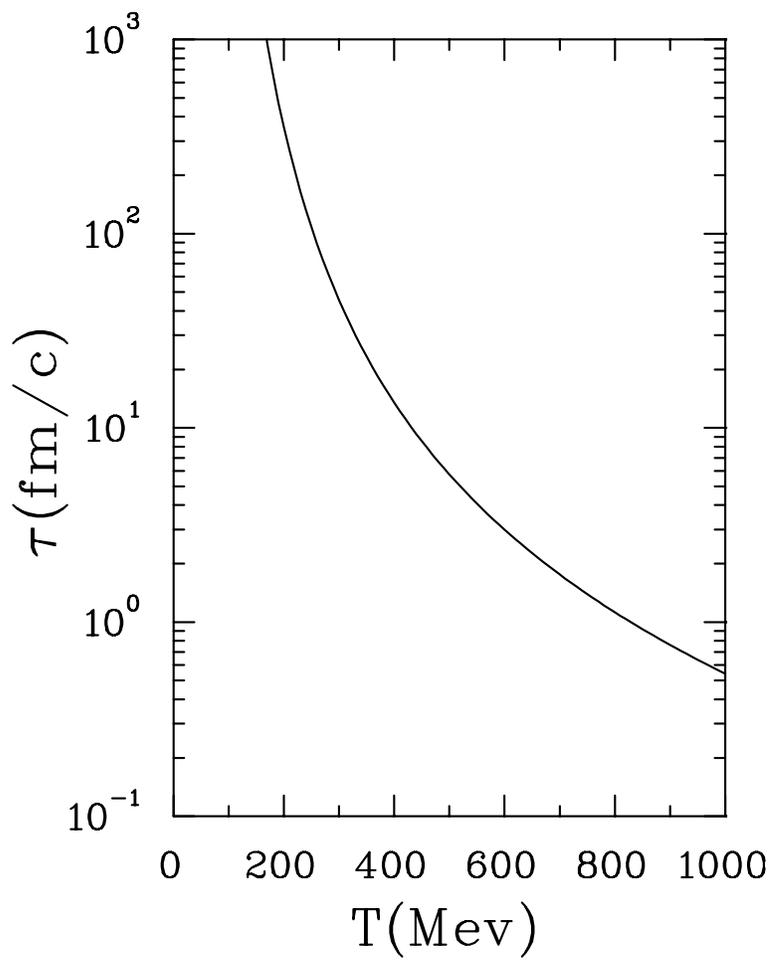}}
\caption{$\Upsilon$ meson dissociation time in a quark-gluon plasma as
function of temperature.}
\end{figure}

\end{document}